\documentclass[12pt]{iopart}

\usepackage{graphicx}
\usepackage{color}
\usepackage{iopams}  

\usepackage{amssymb}
\usepackage{graphics}

\def\pmb#1{\setbox0=\hbox{#1} 
\kern-.020em\copy0\kern-\wd0 
\kern-.04em\copy0\kern-\wd0 
\kern-.020em\raise.0383em\box0} 
\def\SIG{\pmb{$\sigma$}}

\begin{document}

\title[3D dislocation dynamics of LiF micropillar compression at 300K and 600K]{A 3D dislocation dynamics analysis of the size effect on the strength of [111] LiF micropillars at 300K and 600K}

\author{Hyung-Jun Chang$^{1}$, Javier Segurado$^{1, 2}$, Jon M. Molina-Aldaregu{\'\i}a$^{1}$, Rafael Soler$^{1}$, Javier LLorca$^{1, 2, }$}
\address{$^1$ IMDEA Materials Institute, C/ Eric Kandel 2, 28906 - Getafe, Madrid,
  Spain.}
\address{$^2$ Department of Materials Science, Polytechnic University of
  Madrid, 28040 - Madrid, Spain.}
\ead{javier.llorca@imdea.org}

\vspace{10pt}

\begin{indented}
\item[]October 2015
\end{indented}

\begin{abstract}
The mechanical behavior in compression of [111] LiF micropillars with diameters in the range 0.5 $\mu$m to 2.0 $\mu$m was analyzed by means of discrete dislocation dynamics at ambient and elevated temperature. The dislocation velocity was obtained from the Peach-Koehler force acting on the dislocation segments from a thermally-activated model that accounted for the influence of temperature on the lattice resistance. A size effect of the type "smaller is stronger" was predicted by the simulations, which was in quantitative agreement with previous experimental results by the authors \cite{SWC14}.  The contribution of the different physical deformation mechanisms to the size effect (namely, nucleation of dislocations, dislocation exhaustion  and forest hardening) could be ascertained from the simulations and the dominant deformation mode could be assessed as a function of the specimen size and temperature. These results shed light into the complex interaction among size, lattice resistance and dislocation mobility in the mechanical behavior of $\mu$m-sized single crystals.

\end{abstract}

\vspace{2pc}

\noindent Published in Modelling and Simulation in Materials Science and Engineering, {\bf 24} (2016) 035009


\maketitle

\section{Introduction}

It is now well established that crystals exhibit strong size effects in strength at the micron and sub-micron scale, a phenomenon that is commonly referred to as ''smaller is stronger''. This effect has been extensively studied in recent years at room temperature by compressing single-crystal micropillars with diameters in the range 0.5 to 10 $\mu$m \cite{USD09, D09, GD11} and recent studies have covered smaller and larger diameters \cite{DUP05, EUS13, KZS12, SMS07}. These experiments have typically shown that the flow stress of the micropillars increases with decreasing micropillar diameter, albeit the magnitude of the size effect depends strongly on the material system, and more specifically, on the absolute value of the bulk yield strength. For instance, pure metals with the face centered cubic structure \cite{USF04, GON05, VL06} have a virtually negligible bulk yield strength (a few tens of MPa) and display the strongest size effects. In contrast, body centered cubic (BCC) metals possess a substantial bulk yield strength and typically show more moderate size effects \cite{GWC08, SCF09} that actually scale with the critical temperature of the material \cite{SKC09}. And size effects are almost negligible in the case of strong covalent compounds, like GaAs \cite{MWM07} and Si \cite{MWB07}. More recent studies have been focused on the effect of forest hardening and Peierls stress on the size effect \cite{EUS13, SKY13, E15}.

In the absence of solid solution and precipitation hardening, the bulk yield strength is controlled by lattice resistance and the contribution of forest hardening. Considering that the initial dislocation density in all the examples discussed above is of the same order of magnitude, several studies \cite{KC11, SMS12} have proposed that the differences in lattice resistance were responsible for the different size effect in the flow stress in pure metals and compounds. This has been nicely shown in the case of ionic compounds, like MgO \cite{KC11} or LiF \cite{NDU08, SMS12, ZS13}, that display a different size effects depending on crystal orientation. When these compounds are compressed in the [001] direction, slip takes place on the  \{110\}$<$1-10$>$ soft slip systems, giving rise to very pronounced size effects. However, size effects are limited during compression in the [111] direction, where slip occurs on the \{100\}$<$110$>$ hard slip systems. Note that recent experimental and dislocation dynamic studies on Mg also show a significant influence of orientation on the size effect \cite{YMS11, BR13, AFE15}.

The role played by the lattice resistance on the size effect has recently been confirmed by means of high temperature compression of [111] LiF micropillars, with diameters in the range  1 to 5 $\mu$m with an initial dislocation density of 25 $\mu$m$^{-2}$ \cite{SWC14}. It was found that the flow strength was independent of the micropillar diameter at ambient temperature, but a strong size effect developed at 250$^\circ$C. Similar results have also been reported in bcc Mo \cite{SFA13}. In the case of [111] LiF, the different contributions to the flow stress of the micropillars, namely lattice resistance, forest hardening, and the size-dependent contribution were rigorously accounted for as a function of both temperature and micropillar diameter by means of an analytical model. It was demonstrated that the size effect observed during micropillar compression comes about as a result of the relative weights of the size independent (lattice resistance plus forest hardening) and size dependent contributions to the strength. At room temperature, the former dominated, and no size effect was found. Nevertheless, both contributions were of the same order at 250$^\circ$C for the micropillar diameters under study, leading to a strong size effect.

The analytical model in \cite{SWC14} presented, however, several limitations. Firstly, it was able to capture the relative weights of the size independent (lattice hardening plus forest hardening) and the size dependent terms on the initial yield stress as a function of temperature, but could not account for the effect of dislocation velocity, which is also expected to vary significantly with temperature. Secondly, the size-dependent contribution was assumed to be dominated by the operation of single arm dislocation sources. This is a reasonable hypothesis for the initial yield stress if the micropillars are not initially dislocation free \cite{PRD07}, but it is also expected that  dislocation-dislocation and dislocation-surface interactions (that control the initial strain hardening) will also be strongly dependent on the dislocation mobility and, hence, on temperature. Thus dislocation multiplication and dislocation exhaustion phenomena \cite{GD11} can be altered with temperature as a function of the micropillar diameter, and this could not be accounted for in the analytical model.

This paper tries to overcome these limitations by analyzing the mechanical response of [111]  LiF micropillars by means of discrete dislocation dynamics (DDD). The effect of temperature is included in the simulations by means of  a thermally-activated model for the dislocation mobility that accounts for the influence of temperature on the lattice resistance. A size effect of the type "smaller is stronger" was predicted by the simulations, which were in quantitative agreement with the previous experimental results by the authors \cite{SWC14}. Moreover, the contribution of the different physical deformation mechanisms to the size effect (namely, nucleation of dislocations, dislocation exhaustion and forest hardening) could be ascertained from the simulations and the dominant deformation mode could be assessed as a function of the specimen size and temperature.

\section {3D dislocation dynamics framework}

Compression of LiF micropillars was studied by means of 3D DDD. Simulations were carried out using the code K-TRIDIS \cite{FVC97}. The volume under study is discretized by defining a cubic lattice. The cube edge is equal to 5 times the Burgers vector (5$b$ = 2.013 nm). The dislocation lines are defined within that lattice and begin and end at the lattice points. Hence, the minimum length of both edge and screw segments is 2.847 nm (= 5$\sqrt{2}b$). The maximum segment length is limited to 600$b$ and any segment longer than this is further subdivided.

The dislocation line is represented by a succession of pure edge and screw segments. The line vector of the edge segment is perpendicular to the Burgers vector $\mathbf{b}$ while the line vector of the screw segments is parallel to $\mathbf{b}$. Every segment glides along a direction $\mathbf{g}$ perpendicular to its line vector. The shear stress component of the Peach-Koehler force acting on the dislocation segment $i$, $\tau^{pk}_i$,  is given by

\begin{equation} 
 \tau^{pk}_i = \left( \Big[ \mathbf{b} \cdot \big( \hat{ \SIG}_i+ \sum_{j \ne i} \tilde{\SIG}_i^j \big) \Big]\times \mathbf{l} \right) \cdot \mathbf{g} + \tau^{lt}
\label{eq1}
\end{equation}

\noindent where $\mathbf{l}$ is a unit vector parallel to the dislocation line \cite{VFG98}.
$\hat{\SIG}_i$ stands for the contribution of the applied stress at the position of the segment $i$, which also includes the contribution of the image forces on the dislocations due to the crystal boundaries. They are computed  by solving a linear elastic boundary value problem using the finite-element method with the appropriate boundary conditions, following the superposition strategy introduced by Van der Giessen and Needleman \cite{VN95} and extended to 3D in \cite{WFVN02}. The term $\tilde{\SIG}_i^j$  stands for the stress contribution of the $j$th dislocation segment and it is computed analytically from the expressions for the stress field induced by a dislocation segment on an infinite, elastic and isotropic continuum according to \cite{W67, L83}. Finally $\tau^{lt}$ is the line tension, which is calculated from the local radius of curvature of the dislocation line according to 

\begin{equation}
\label{eq_tault}
\tau^{lt}= \frac{\alpha \mu b}{R} 
\end{equation}

\noindent where $R$ is the radius of circle generated from center points of the segment and its two neighbor segments, $\mu$ the shear modulus of the material and  $\alpha$ = 0.63 \cite{VFG98}. 

The velocity of each dislocation segment along the slip direction was computed in each time increment as indicated below and the equations of motion of the dislocations were solved using an explicit incremental algorithm.
 
\begin{equation}
\label{eq_velinc}
u^{t + \Delta t}= u^{t} +v_t \Delta t
\end{equation}

\noindent where the velocity in the current time step ($v_t$) is obtained from the thermal activation model described in the next paragraph. The short range interactions between dislocations (annihilation, recombination, junction formation, etc.) were taken into account after each time increment. Charge effects were not considered and only the hard slip systems have been accounted explicitly for in the simulations. Due to the orientation of the sample, most of mobile dislocations belong to the hard slip systems because the Schmid factor of the soft slip systems is zero. Thus, dislocation segments in the soft slip system would remain, on average, immobile and will not contribute to the plastic deformation. Their effect on the movement of the dislocations in the hard slip system was indirectly accounted for in the dislocation mobility law, equation (4), which provides the probability of the mobile dislocations to overcome the barriers for a given temperature and stress. The parameters in equation (4) were obtained from the micropillar compression tests at different temperatures in samples containing an initial dislocation density in the soft slip system \cite{SWC14}. These dislocations in the soft slip system acted as barriers to the dislocation motion and their density  did not change during deformation.

Cross slip was not included in the model because cross-slip in the LiF structure can occur only between the soft and hard slip systems. If a screw dislocation segment crosses from a hard to a soft slip system, this segment would remain immobile and will not contribute to the plastic deformation because of the zero Schmid factor. In addition, it can be assumed that the probability of cross slip in this case is very small because it depends, among other factors, on the Schmid stress on both the original and the cross slip plane \cite{KYC14, HRU15} and the latter is zero in the soft planes. In the absence of more specific studies about cross slip in LiF structures and due to the reasons previously presented, the current DDD model did not take into account a cross slip. More details about the 3D DDD model and the coupling with the finite element solver can be found in \cite{VFG98, CHANG}.
 				   			
Assuming a standard approach for a stress-activated process, the velocity of each dislocation segment along the slip direction, $v$, can be determined from the shear component of the Peach-Koehler force acting on the dislocation segment, $\tau^{pk}$ according to

\begin{equation}
\label{eq_disvel}
v = \nu b \left[ \exp\Bigg(-\frac{(\tau_p-\tau^{pk})V}{kT}\Bigg)- \exp\Bigg(-\frac{(\tau_p+\tau^{pk})V}{kT}\Bigg) \right] 
\end{equation}

\noindent   where $\nu$ is the attempt frequency, $\tau_p$ the Peierls stress, $V$ the activation volume, $T$ the absolute temperature, and $k$ the Boltzmann's constant. Although $\nu$ should be dependent on the length of dislocation segments and temperature \cite{ACD11}, it was considered constant and equal to 8.0 10$^{11}$ s$^{-1}$ for simplicity.  
$V$ = 11.41$b^3$ (where $b$ = 0.285 nm) was obtained for LiF subjected to plastic deformation along the hard \{100\}$<$110$>$ slip system by means mechanical tests at different strain rates at ambient temperature \cite{SWC14}. Similarly, the Peierls stress $\tau_p$ = 530 MPa was obtained from experimental results of the critical resolver shear stress of bulk LiF at different temperatures \cite{SWC14}.

\begin{figure}[h!]
\centering
\includegraphics[scale=1]{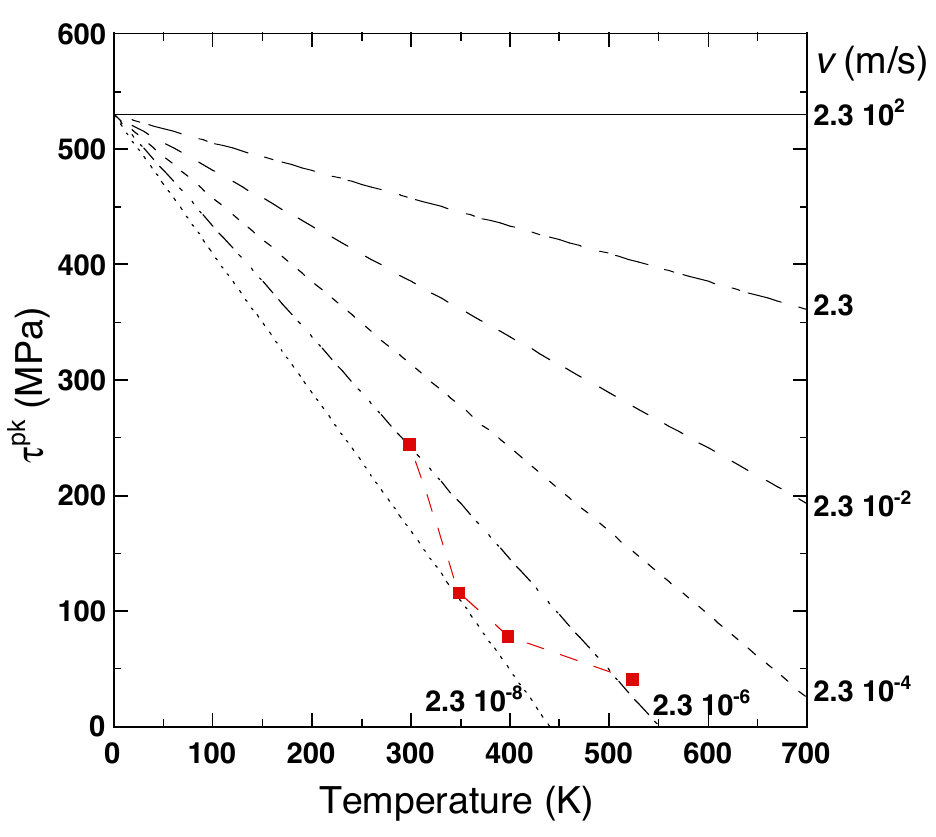}
\caption{Loci of constant dislocation velocity $v$ as a function of the critical resolved shear stress, $\tau^{pk}$ and the temperature  $T$ for LiF deformed along the hard {100}$<$110$>$ slip system. Solid symbols correspond to the micropillar compression tests at different temperatures reported in \cite{SWC14}.}
\label{fig:disvel} 
\end{figure}

If  $\tau_{pk} V >> kT$, equation (\ref{eq_disvel}) can be approximated by

\begin{equation}
\label{eq_disvel2}
v \approx  \nu b \exp\Bigg(-\frac{(\tau_p-\tau^{pk})V}{kT}\Bigg)
\end{equation}

\noindent and the straight lines in Fig. \ref{fig:disvel} stand for the loci of constant dislocation velocity for different values of the applied critical resolved shear stress and temperature. The solid symbols in Fig. \ref{fig:disvel} correspond to the micropillar compression tests reported in \cite{SWC14} at different temperatures and the average dislocation velocity was in the range  2.3 10$^{-6}$ to 2.3 10$^{-8}$ m/s. According to Orowan, the shear strain rate $\dot\gamma$, the mobile dislocation density $\rho_m$, the Burgers vector $b$ and the average dislocation velocity $v$ are related by

\begin{equation}
\label{eq_orowan}
\dot\gamma = \rho_m b v
\end{equation}

\noindent  where $\rho_m \approx $ 10$^{13}$ m$^{-2}$ \cite{SWC14} and this expression leads to a shear strain rate $\dot\gamma \approx$ 6.6 10$^{-3}$ - 6.6 10$^{-5}$ s$^{-1}$, which is very close to the experimental one \cite{SWC14}, as expected.

Unfortunately, DDD simulations have to be carried out at much higher strain rates because of small time step required for the stable integration of the equations of motion. Assuming an applied strain rate of 2000 s$^{-1}$, equation (\ref{eq_orowan}) imposes an average dislocation velocity close to 1 m/s. The stresses necessary to move dislocations at this velocity are much higher than those measured in the micropillar compression tests at under quasi-static conditions, according to equation (\ref{eq_disvel}). In this high stress regime, the influence of the temperature on the dislocation velocity is very small, leading to very different deformation mechanisms. Thus,  it is necessary to bridge this gap in order to obtain useful information from the DDD simulations, and this was achieved by increasing artificially the dislocation velocity according to 

\begin{equation}
\label{eq_disvel3}
v \approx   W \nu b \exp\Bigg(-\frac{(\tau_p-\tau^{pk})V}{kT}\Bigg)
\end{equation}

\noindent where $W$ = 10$^{9}$ is a factor that brings the dislocation dynamics simulations close to the experimental results. Besides, the factor approximately compensates the difference of loading rate between DDD simulation (2000 s$^{-1}$) and experiments (2.0 10$^{-3}$ s$^{-1}$). In particular, the applied stress and the dependence of the dislocation mobility with temperature in the DDD simulations are similar to the experimental ones. A cut-off in the dislocation velocity of 2000 m/s was introduced in the simulations to avoid numerical instabilities. 

\subsection {Numerical model}
3D DDD simulations were carried out in cylindrical LiF single crystals oriented in the $<$111$>$ direction. Micropillars with diameters $D$ of 0.5, 0.67, 1, 1.5 and 2 $\mu$m were analyzed to assess the influence of micropillar size on the strength and the deformation micromechanisms. The aspect ratio $L/D$ of the micropillars was constant and equal to 2. The elastic behavior of the micropillars was characterized by the shear modulus $\mu$ (63.6 GPa) and the Poisson's ratio (0.216)  of the LiF crystal.  Plastic deformation occurred by dislocation slip along three slip systems, [101](010), [011](100) and [110](001), corresponding to the \{100\} $<$110$ >$ slip system. Their Schmid factor during compression along the $<$111$>$ direction is 0.47. The slip planes corresponding to the \{110\} $<$110$>$ slip system were not included in the model because the Schmid factor is zero and the experimental evidence showed that plastic deformation took place in the  \{100\} $<$110$>$ slip system at all temperatures \cite{SWC14, SMS12a}.

The displacements of the bottom surface of the cylinder were fully constrained while the top surface was deformed along the cylinder axis at a constant strain rate of $\dot\epsilon$ = -2000 s$^{-1}$. The lateral surfaces of the cylinder were stress free and the dislocations were free to exit the crystal through the lateral surfaces. 

\begin{figure}
\centering
\includegraphics[width=11cm]{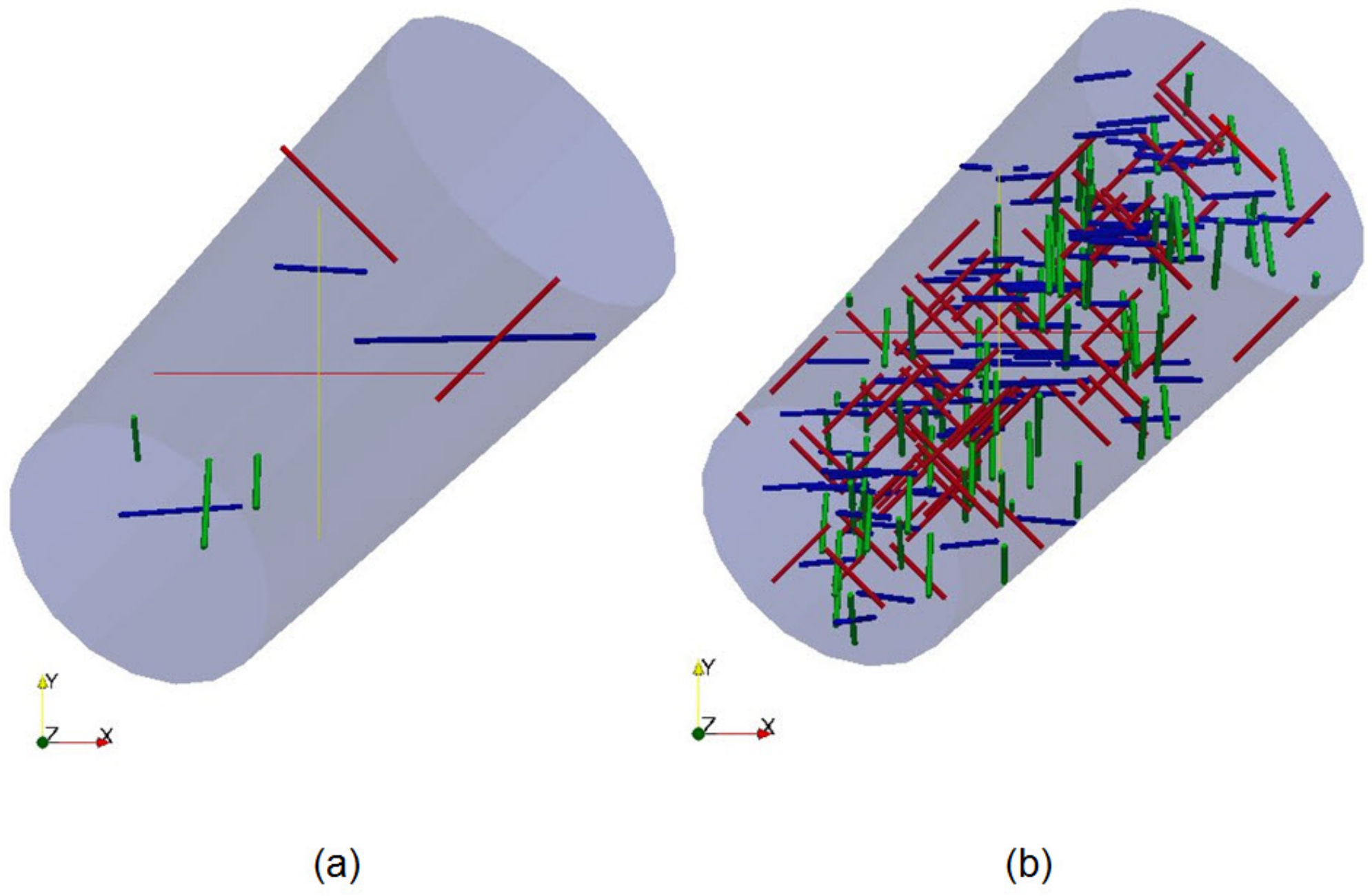}
\includegraphics[scale=0.7]{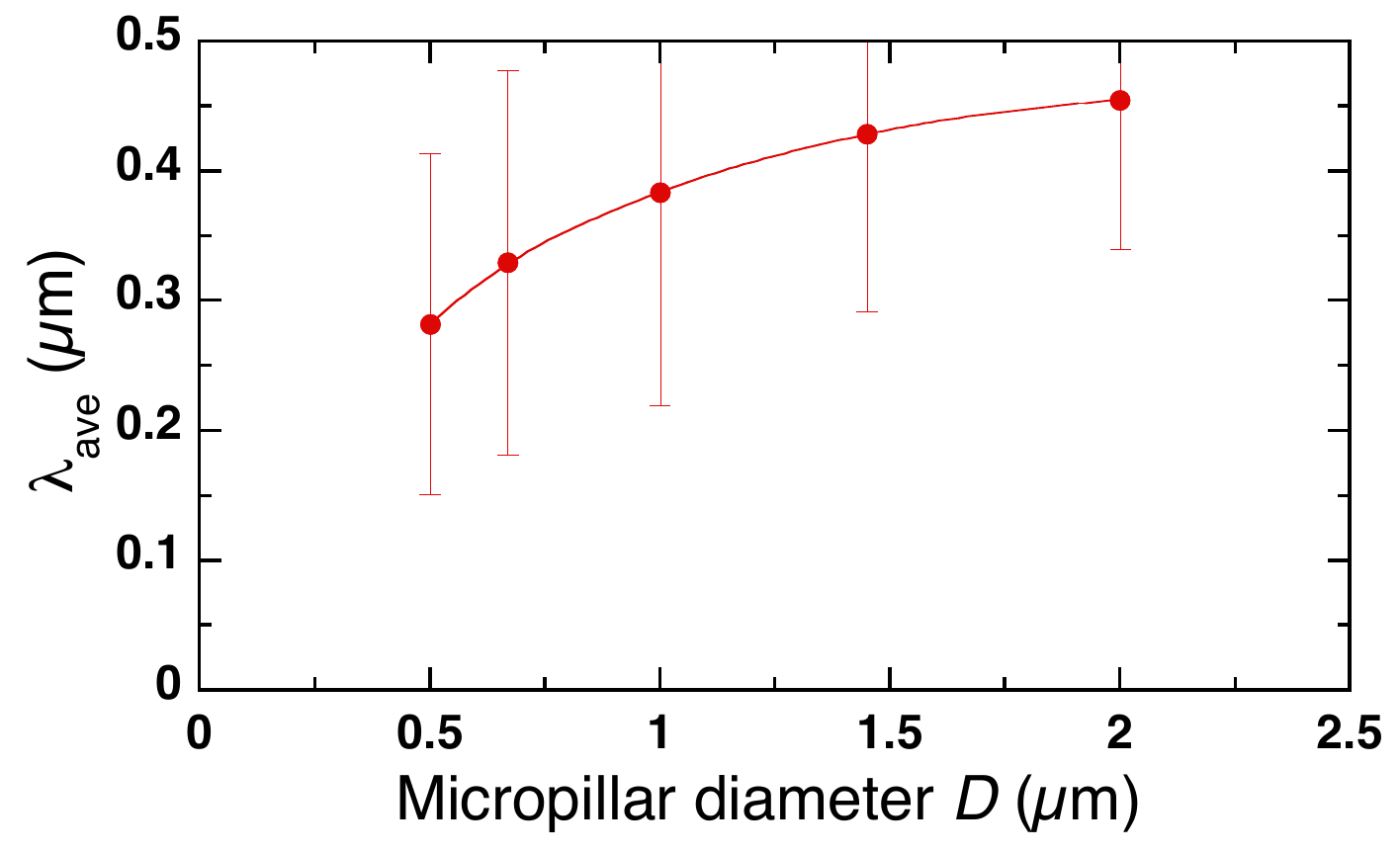}
\caption{ (a) Initial distribution of Frank-Read sources in [111] LiF micropillar with $D$ = 0.5 $\mu$m. (b) {\it Idem} with $D$ = 2.0 $\mu$m. Different colors stand for different slip systems. (c) Average length (and standard deviation) of the length of the Frank-Read dislocation sources in micropillars with different diameter. The data for each micropillar diameter correspond to three different realizations of the Frank-read sources.}
\label{fig:ini} 
\end{figure}

A random distribution of Frank-Read sources was included in the three slip systems in each micropillar, Figs. \ref{fig:ini}(a) and (b). The initial dislocation density $\rho$ due to the Frank-Read sources was 9 $\mu$m$^{-2}$, similar to the one measured in the LiF micropillars \cite{SWC14}. To this end, Frank-Read sources of 0.5 $\mu$m in length were randomly placed in the three slip systems. Many Frank-Read sources, particularly in the smaller micropillars, cut the micropillar surface  and the actual length distribution of the initial Frank-read sources depended on the micropillar diameter. Thus, the average length of the Frank-Read sources, $\lambda_{ave}$,  increased with micropillar diameter, as shown in Fig. \ref{fig:ini}(c). Note that the variation of average source length will directly introduce an athermal size effect \cite{RDP08, EWG09}

\section{Results and Discussion}

3D DDD simulations were carried out in micropillars of different diameter, from 0.5 $\mu$ to 2 $\mu$m. For each micropillar size, simulations were carried out with three different distributions of the initial Frank-Read sources at 300K and 600K. The effect of the temperature in the DDD was included through the mobility law for screw segment (Eq. \ref{eq_disvel3}) and it was assumed that the velocity of the edge dislocation segments was ten times higher than that of the screw segments at 300K while screw and edge dislocations have the same velocity at 600K, according to the experimental evidence in LiF \cite{JG59, ACD11}.

\begin{figure}
\includegraphics[scale=0.78]{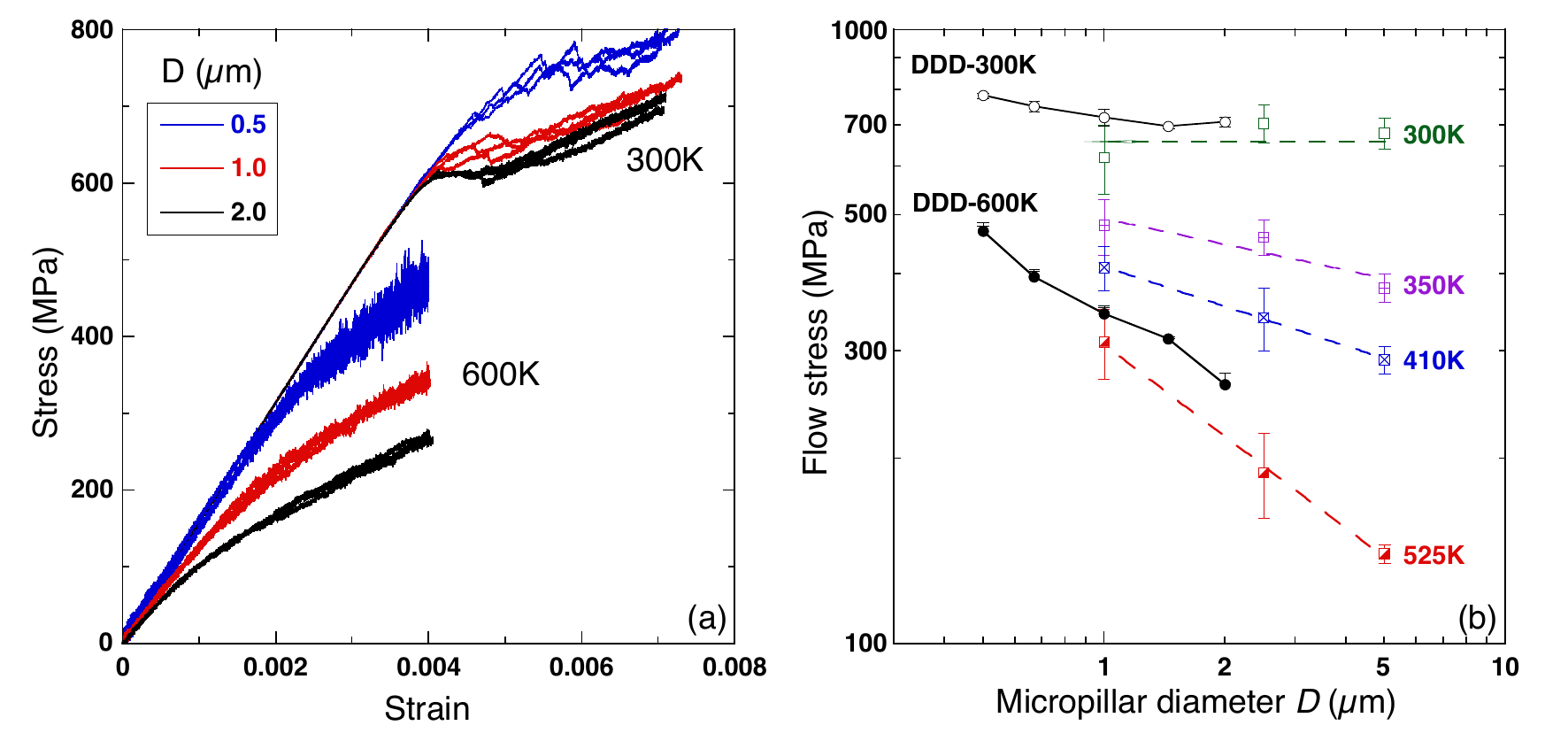}
\caption{(a) Stress-strain curves in compression of micropillars with $D$ = 0.5, 1.0 and 2.0 $\mu$m at 300K and 600K obtained by means of DDD simulations. (b) Experimental and DDD results of the flow stress of [111] LiF micropillars as a function of the micropillar diameter and test temperature. Experimental results are given in \cite{SWC14}.}
\label{fig:S-E} 
\end{figure}

The compressive stress-strain curves of the micropillars with $D$ = 0.5, 1.0 and 2.0 $\mu$m in diameter at 300K and 600K are plotted in Fig. \ref{fig:S-E}(a). Three curves are plotted for each micropillar size and temperature. The fluctuations in the stress-strain curves (particularly for small micropillars at high temperature) are numerical artifacts induced by the constant strain rate loading condition when some dislocations are leaving the micropillar. Strong image forces develop when the dislocation segment approaches the free surface but they disappear suddenly as the segment exits the pillar. This leads to lead to large fluctuations in the force to attain mechanical equilibrium, which are particularly noticeable for small  micropillars at high temperature. The curves corresponding to the micropillars of 0.67 $\mu$m and 1.5 $\mu$m in diameter followed the same trends and were not plotted in Fig. \ref{fig:S-E}(a) for the sake of clarity. The scatter among simulations with different realizations of Frank-Read sources was very limited for all micropillar diameters at both temperatures and a size effect of the type "smaller is stronger" is clearly predicted by the DDD simulations. This behavior has already been found by other authors in metallic FCC micropillars using DDD \cite{BS06, SWG08, TSE07, AZB10, ZBL11}, and it has been attributed to different mechanisms. The size effect in the critical stress for the onset of yielding could be reasonably predicted by the activation of truncated single-arm dislocation sources, following the postulates of \cite{PRD07}, because the maximum length of the Frank-Read sources (which controls the activation of plastic slip) decreases together with the micropillar diameter. Further plastic strain led to the development of two dominant hardening mechanisms as a function the micropillar diameter. In  small micropillars ($D \lessapprox$ 0.5 $ \mu$m), dislocation exhaustion, owing to the escape of mobile dislocations through the lateral surfaces, became the dominant hardening mechanism. Deformation of larger micropillars ($D \gtrapprox$ 1.0 $ \mu$m) will be dominated  by forest hardening because of the development of dislocation reactions which hinder the mobility of the dislocations. As noted by \cite{ZBL11}, the frontiers between both types of hardening are highly sensitive to the initial dislocation density and to the internal dislocation structures. Nevertheless, the influence of the lattice resistance and of the dislocation mobility was only accounted for by  \cite{AFE15}.

While these previous DDD simulations of micropillar compression were in qualitative agreement with experimental results in the literature, quantitative comparisons were difficult due to the large differences in strain rate between the simulations and experiments.  Only in a few cases it was possible to carry out direct comparisons between experiments and DDD simulations with similar strain rates \cite{RDP08, E15}. This limitation is overcome in our simulations owing to the dislocation mobility law (eq. \ref{eq_disvel3}) and the DDD predictions of the flow stress for micropillars of different diameter were in quantitative agreement with the experimental results at ambient and elevated temperature, Fig. \ref{fig:S-E}(b). Moreover, the DDD simulations predicted a negligible size effect at 300K and a strong size effect at 600K, in excellent agreement with experiments. The magnitude of the size effect at 600K was also very close to the experimental results. This quantitative agreement is somehow surprising because the experimental values of the flow stress in Fig. \ref{fig:S-E}(b) were measured at an applied strain of $\approx$ 5\%, where the DDD simulations were stopped at applied strains of 0.4\% (600K) and 0.7\% (300K) and the flow stresses in Fig. \ref{fig:S-E}(b) correspond to these strains. It should be noted, however, that the experimental flow stress of the micropillars tested at high temperature was almost constant for applied strains beyond 1\% \cite{SWC14} and this explains the accuracy in the quantitative predictions of the size effect.

\begin{figure}[h!]
\centering
\includegraphics[scale=1]{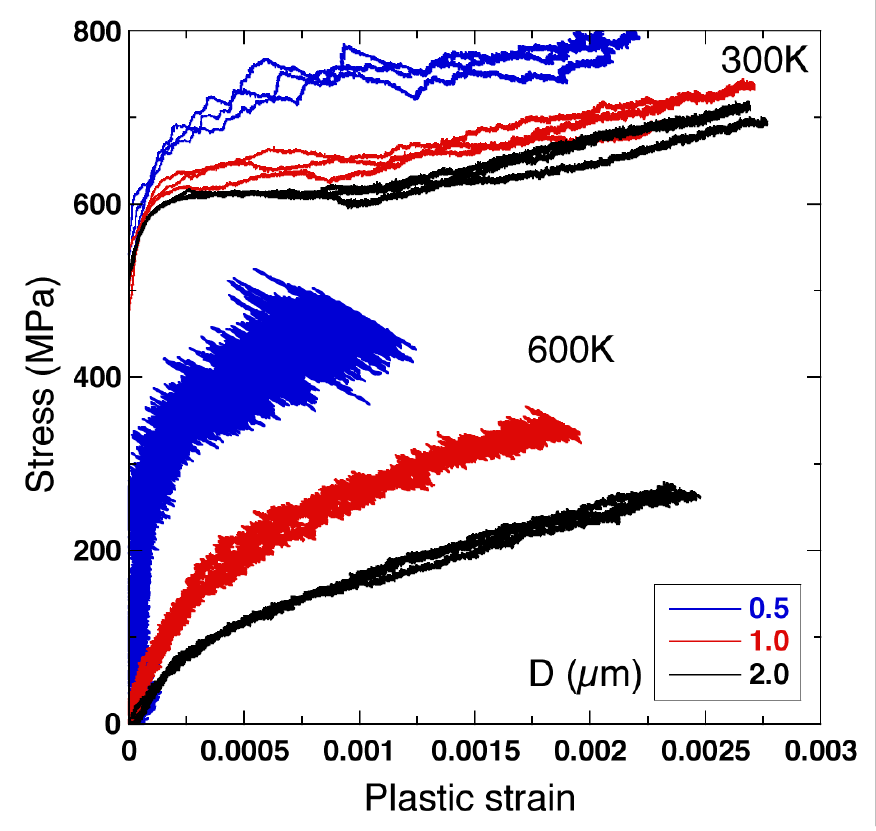}
\caption{Stress {\it vs.} plastic strain curves in compression of micropillars with $D$ = 0.5, 1.0 and 2.0 $\mu$m at 300K and 600K obtained by means of DDD simulations.}
\label{fig:S-Ep} 
\end{figure}

The flow stress from the DDD simulations is plotted in Fig. \ref{fig:S-Ep} as a function of the plastic strain for micropillars of different diameter tested at 300K and 600K. In this way, the different contributions to the hardening of the micropillars can be better understood. At 300K, the size effect develops very quickly during the elasto-plastic transition and reaches a maximum value at $\epsilon_p$ = 0.1\%. Further plastic deformation led to strain hardening but the hardening rate was independent of the micropillar diameter and the size effect did not change with the plastic strain. In the case of the simulations of micropillar compression at 600K, the size effect also developed from the first stages of plastic deformation but increased with the plastic strain. In addition, a significant fraction of the total applied strain (0.4\%) in the smallest micropillar was accommodated by elastic deformation ($\approx$ 0.25\%).

The initial size effect at 300K and 600K predicted by the DDD can be rationalized in terms of the activation of the Frank-Read sources. The contribution of the differences in the length of the Frank-Read sources to the size effect can be estimated according to the line tension model as

\begin{equation}
\label{eq:FR}
\sigma_{FR}=\frac{2 \alpha \mu b}{M\lambda_{ave}}
\end{equation} 

\noindent where the $\sigma_{FR}$ is the stress necessary to promote dislocation nucleation a Frank-Read source of length $\lambda_{ave}$, $M$ (=0.47) is the Schmid factor for the slip system and $\alpha$ = 0.63 \cite{VFG98}. As the variation of $\mu$ and $b$ with temperature is negligible, this  contribution is independent of the temperature. According to the values of the average length of the Frank-Read sources in Fig. \ref{fig:ini}(c), the compressive stress necessary to promote dislocation motion should be 172 MPa for the micropillars with $D$ = 0.5 $\mu$m and 107 MPa for $D$ = 2.0 $\mu$m. This difference of around 70 MPa between the largest and the smallest micropillar is in good agreement with the stress-strain curves at the onset of plastic deformation at 300K and 600K (Fig. \ref{fig:S-E})(a)

\begin{figure}
\centering
\includegraphics[scale=0.8]{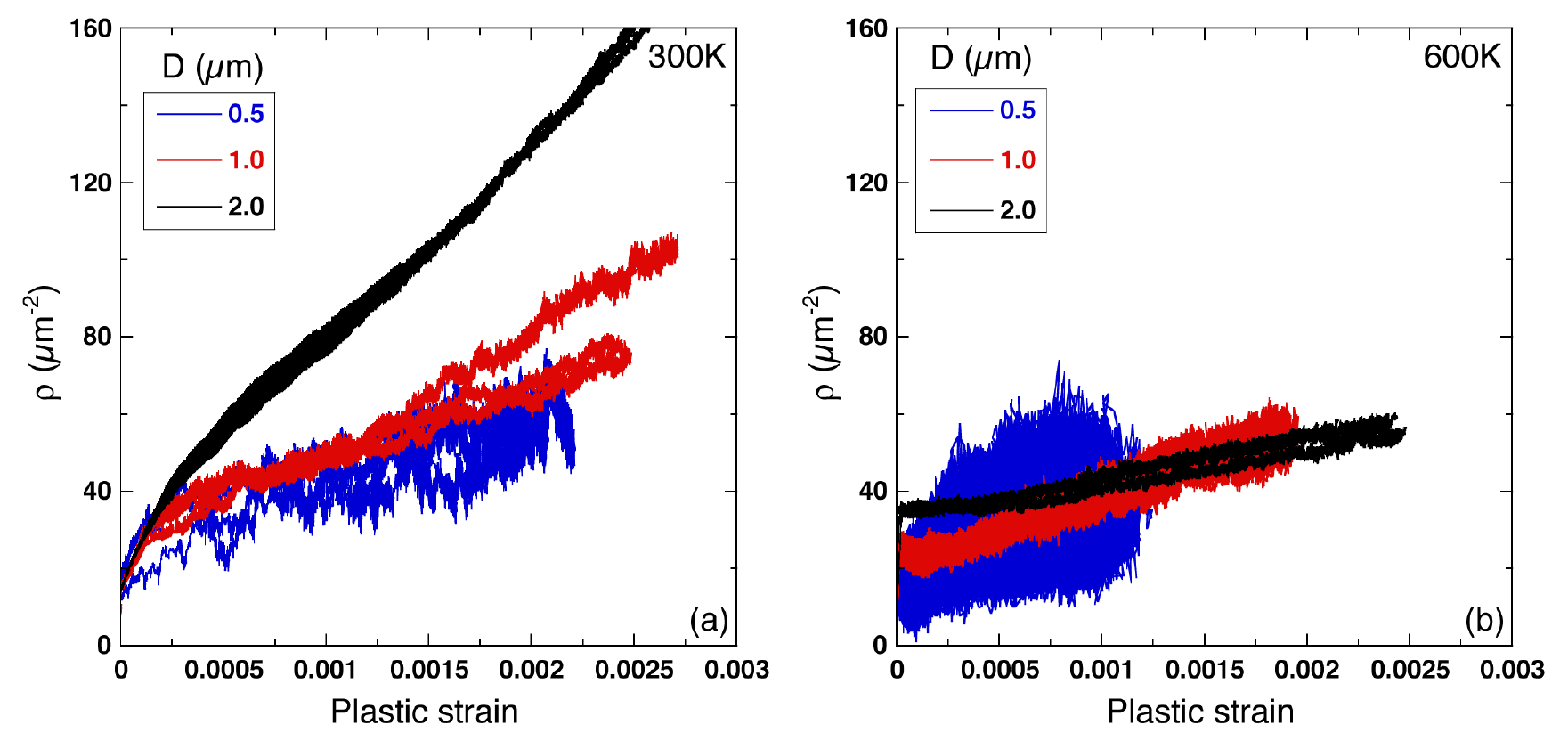}
\caption{DDD results of the evolution of the dislocation density, $\rho$, as a function of the applied plastic strain during compression of [111] LiF micropillars.   (a) 300K  and (b) 600K. The data for each micropillar diameter correspond to three different realizations of the Frank-read sources.}
\label{fig:DD-Ep} 
\end{figure}

\begin{figure}
\centering
\includegraphics[scale=0.8]{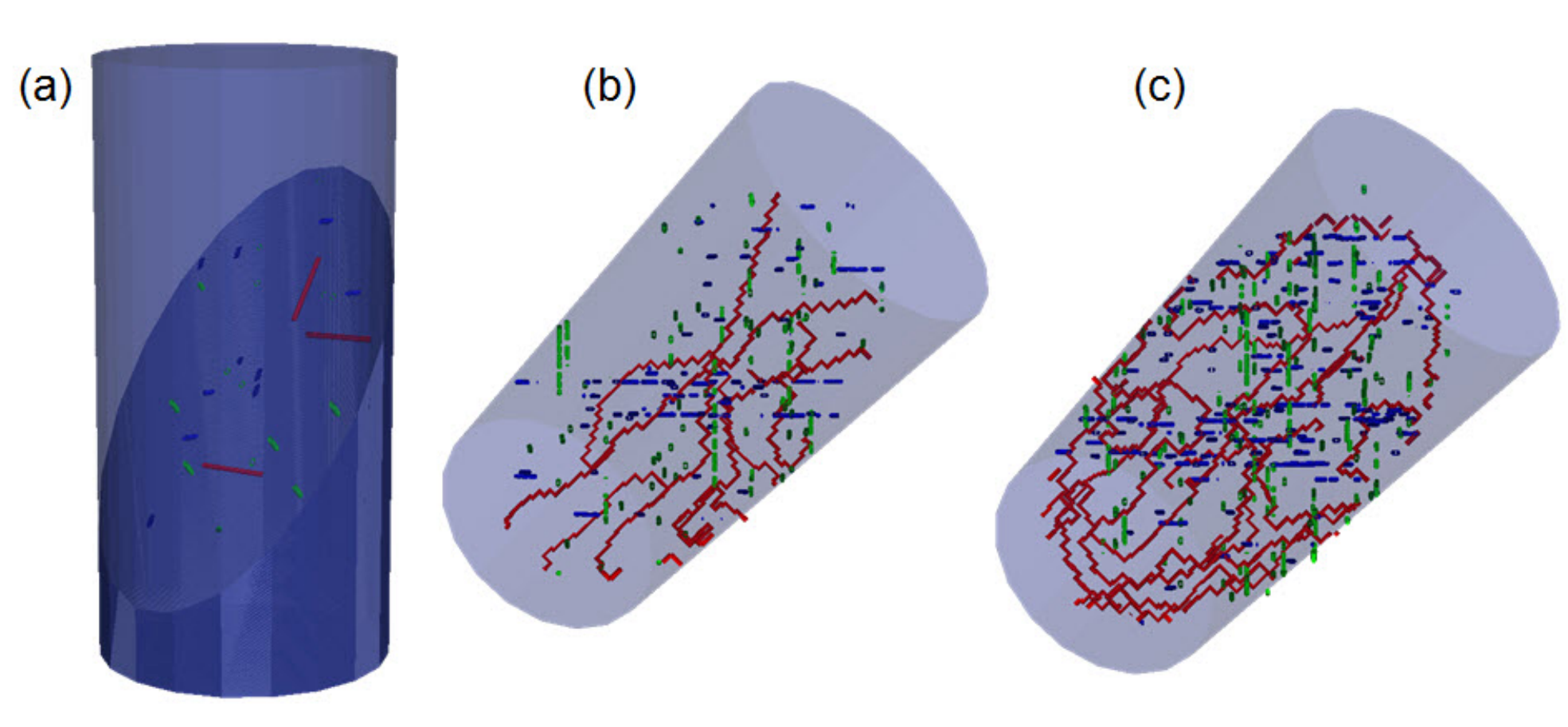}
\caption{Dislocation evolution in one section of the pillar along the [110] plane. The colors stand for the different slip systems: red corresponds to the dislocations in slip system [110] (001) which are contained in the shown section, while blue and green show the dislocations belonging to the slip systems [101](010) and [011] (100), respectively $D$ = 2.0 $\mu$m. $T$ = 300K.   (a) Initial distribution of Frank-Read sources in the slice.  (b) $\epsilon_p$ = 0.1\%. (c) $\epsilon_p$ = 0.2\%.}
\label{fig:meso} 
\end{figure}

In order to understand the dominant deformation mechanisms after the onset of yielding, it is interesting to analyze the evolution of the dislocation density with the plastic strain (Fig. \ref{fig:DD-Ep}).  The dislocation density increased with the plastic strain and with the pillar diameter in the micropillars compressed at 300K, Fig. \ref{fig:DD-Ep}(a). The large dislocation density attained at the end of the simulations in the case of the micropillar with $D$ = 2 $\mu$m  is the result of the development of tangled dislocation structures due to the development of multiple dislocation-dislocation interactions. This is illustrated in Fig. \ref{fig:meso}, which shows the evolution  of the dislocation structures in one slice of the pillar along the [110] plane. The colors stand for the different slip systems: red corresponds to the dislocations in slip system [110] (001) which are contained in the shown section, while blue and green show the dislocations belonging to the slip systems [101](010) and [011] (100), respectively. In these two cases, dislocations are not contained in the shown section and the segments shown in the figure are the intersections between the dislocation line and the [110] plane. The dislocation structure shows the development of ellipsoidal loops, where the loop shape depends on the ratio of velocities between edge and screw segments. The shape of dislocation loops in the section is quite complex because it is intercepted by dislocations on the other slip systems (shown with different colors)
From the initial distribution of Frank-Read sources, a complex dislocation network is formed as the plastic strain $\epsilon_p$ increases. This mesostructure is responsible for the forest hardening observed at 300K  in the stress-strain curves (Fig. \ref{fig:S-Ep}) of the micropillars with large diameter for plastic strains above 0.1\%. 

In the case of the micropillar with $D$ = 0.5 $\mu$m at 300K, the increase in dislocation density with strain was moderate and both the stress {\it vs.} strain and the dislocation density {\it vs.} strain curves presented a serrated shape. This behavior is indicative of  hardening due to dislocation exhaustion: mobile dislocations leave the crystal through the surfaces before they can be stopped by other dislocations. There are not enough mobile dislocations within the crystal to accommodate the imposed strain rate and elastic stresses are build up to activate more dislocation sources. The new mobile dislocations absorb part of the elastic strain, leading to a reduction in the flow stress until they exit the crystal and the whole process is repeated. It is worth noting that the hardening rate due to dislocation forest hardening and to exhaustion hardening was very similar at 300K and thus the size effect becomes independent of the applied strain at 300K. 

The results for the evolution of the dislocation density with plastic strain at 600K, Fig. \ref{fig:DD-Ep}(b), differ, however, from those found at 300K. In particular, the increase in dislocation density is limited regardless of the micropillar diameter. This is indicative that dislocation-dislocation interactions could not develop at high temperature and an equilibrium was attained between the dislocation nucleation and escape rates in the largest micropillar. 
These results are consistent with previous molecular dynamics simulations on the effect of temperature on junction formation and release \cite{HSR14}, which showed that the dipole length (maximum distance not to break junction) and the critical stress for unzipping decreased as the temperature increased. 
As the micropillar diameter decreased, hardening by the dislocation exhaustion mechanism was clearly dominant, and it was again revealed in the jerky behavior of the stress {\it vs.} strain and dislocation density {\it vs.} strain curves. Thus, the size effect increased with the applied strain at high temperature due to dislocation exhaustion.

These DDD simulations show that temperature influences the size effect during micropillar compression of [111] LiF crystals through two different mechanisms related to the variation of the lattice resistance and of the dislocation mobility. The reduction of the lattice resistance with temperature leads to a more noticeable size effect at high temperature, because the athermal contribution of the size effect (due to the average length of the initial distribution of Frank-Read sources) is of the same order as the lattice resistance at high temperature while it becomes of second order at high temperature \cite{SWC14}. This is in agreement with the experimental evidence, which indicates that size effect during micropillar compression of strong solids is almost negligible \cite{MWM07, RMW13, SMS12}

\begin{figure}[h!]
\centering
\includegraphics[scale=01.0]{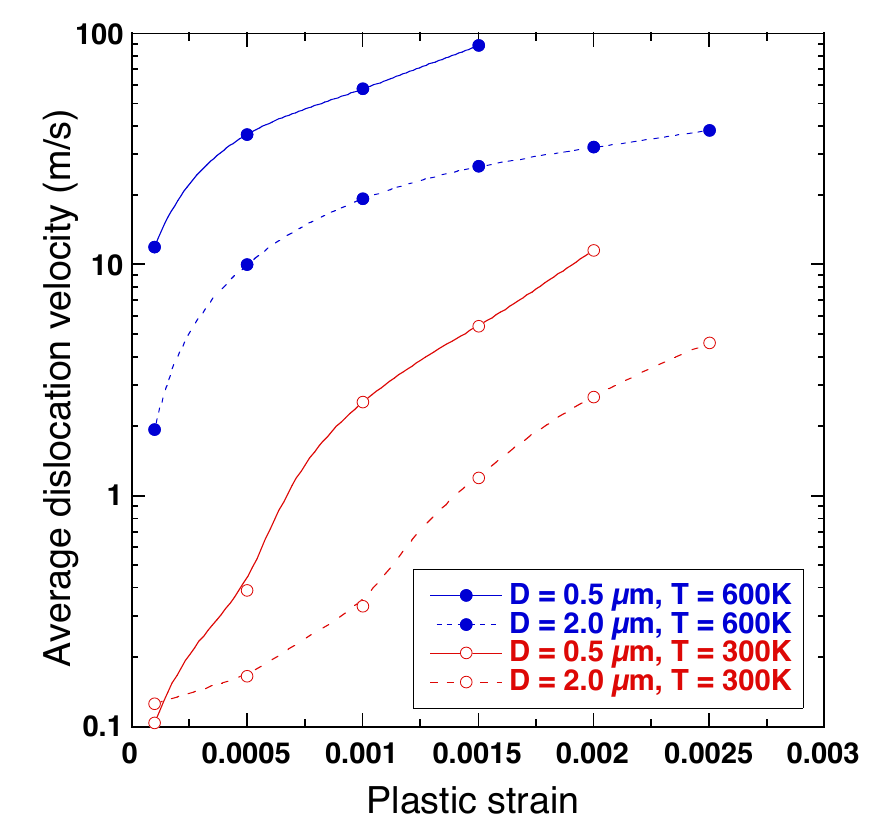}
\caption{Evolution of the average dislocation velocity $\overline v$ as a function of the plastic strain for micropillars of different diameter at 300K and 600K.}
\label{fig:Dv-Ep} 
\end{figure}

Changes in dislocation mobility with temperature also play a critical role on the development of size effects in small solids. Following Orowan's equation, the average dislocation velocity, $\overline v$ can be computed as

\begin{equation}
\label{eq:aveDv}
\overline v = \frac{\dot\epsilon_p}{b \rho}
\end{equation} 

\noindent where the magnitude of $\dot\epsilon_p$ and $\rho$ can be obtained from the DDD simulations \cite{FD14} and the average dislocation velocity in the micropillars with $D$ = 0.5 $\mu$m and 2 $\mu$m at 300K and 600K is plotted in Fig. \ref{fig:Dv-Ep}. These results show that the average dislocation velocity at 600K is over one order of magnitude higher than at 300K for the same sample size. Under these conditions, the dislocations in micropillars tested at high temperature leave the crystal without interacting with other dislocations, and deplete the sample from mobile dislocations, activating the mechanisms of hardening by the dislocation exhaustion. This phenomenon is more dramatic as the micropillar size is reduced and the applied stress is increased, leading to a  large  size effect on the micropillar flow stress that, in addition, increases with the applied strain. The lower mobility of the dislocations at 300K, however, facilitates the development of tangled dislocation structures due to dislocation-dislocation interactions. Dislocations are hindered to leave the crystal and hardening takes place by forest hardening rather than by dislocation exhaustion, particularly in the largest micropillar.

\section{Conclusion}

The mechanical response in compression of [111] LiF micropillars with diameters in the range 0.5 $\mu$m to 2.0 $\mu$m at 300K and 600K was studied by means of DDD simulations. An initial dislocation density of 9 $\mu$m$^{-2}$ was introduced in the single crystals along the three {100}$<$110$>$ slip planes. The Peach-Koehler
force acting on the dislocations was computed from the applied stress, the contribution of all the dislocation segments and the image forces. The dislocation velocity was obtained from the Peach-Koehler force from a thermally-activated model that accounted for the influence of temperature on the lattice resistance. This velocity was scaled by a bridging factor to account for the different strain rate between the DDD simulations (2000 s$^{-1}$) and the experiments in \cite{SWC14} ($\approx$ 10$^{-3}$ s$^{-1}$).

The DDD simulations predicted a size effect in the flow stress of the micropillars of the type " smaller is stronger" and were in excellent quantitative agreement with the experimental results in \cite{SWC14}. Three sources of size effects were identified by the simulations. The initial flow stress was controlled by the activation of Frank-Read sources, whose average length decreased with micropillar diameter. The hardening rate after the onset of yielding depended on the average dislocation velocity.  Dislocations left rapidly the micropillar through the lateral surfaces without interacting with other dislocations at high temperature  (600K) in small micropillars ($D$ = 0.5 $\mu$m). The dislocation density was low and there were not enough mobile dislocations to accommodate the applied strain, leading to elastic hardening by a dislocation exhaustion mechanism. On the contrary, dislocations were trapped by other dislocations at ambient temperature (300K) in the large micropillars ($D$ = 2 $\mu$m), the dislocation density increased with the applied strain and forest hardening was dominant under these conditions. Obviously, the transition from one mechanism to another is dependent on the micropillar size and on the influence of the temperature on the dislocation mobility, which in turn is controlled by the actual mechanism of dislocation slip. The excellent agreement between experiments and simulations shows the potential of DDD to explore the development of size effects in small solids by including the contribution of the different physical mechanisms of strain hardening (nucleation of dislocations from single-arm sources, dislocation exhaustion and forest hardening) during deformation as a function of temperature and strain rate. 

\ack
This investigation was supported by the European Research Council (ERC) under the European UnionÕs Horizon 2020 research and innovation programme (Advanced Grant VIRMETAL, grant agreement No. 669141). In addition, HJC acknowledges the support of Spanish Ministry of Economy and Competitiveness through the Juan de la Cierva fellowship.

\section*{References}

\end{document}